# Influence of Plaque Characteristics on Stent Biomechanical Outcomes – A Case Study on Double Kissing Crush Coronary Stenting

Andrea Colombo[a], Dario Carbonaro[b,c], Mingzi Zhang[a,d], Chi Shen[a], Ankush Kapoor[a], Nigel Jepson[e,f], Claudio Chiastra[b], Susann Beier[a]

[a] Sydney Vascular Modelling Group, School of Mechanical and Manufacturing Engineering, University of New South Wales, Sydney, NSW, Australia

[b] Polito$^{BIO}$Med Lab, Department of Mechanical and Aerospace Engineering, Politecnico di Torino, Turin, Italy

[c] Department of Mechanical, Energy and Management Engineering, University of Calabria, Rende, Italy

[d] Centre for Healthy Futures, Torrens University Australia, Surry Hills, Sydney, Australia

[e] Prince of Wales Clinical School of Medicine, University of New South Wales, Sydney, NSW, Australia

[f] Prince of Wales Hospital, Sydney, NSW, Australia

**Corresponding author:**
Andrea Colombo
School of Mechanical and Manufacturing Engineering, Ainsworth Building (J17), Engineering Rd, University of New South Wales, Kensington, NSW, 2052, Australia
Email: a.colombo@unsw.edu.au
https://orcid.org/0009-0009-6436-6785

**Statements and Declarations**

**Competing interest**
The authors have no competing interests to declare that are relevant to the content of this article.
**Funding**
No funding was received for this work.
**Ethics**
No ethical approval was required for this work.

**Author Contribution Declaration**

**Andrea Colombo:** Conceptualization, Methodology, Investigations, Visualization, Writing - Original Draft, Project administration. **Dario Carbonaro**: Methodology, Writing - Review & Editing, Supervision. **Mingzi Zhang**: Methodology, Writing - Review & Editing. **Chi Shen**: Methodology, Writing - Review & Editing. **Ankush Kapoor**: Methodology, Writing - Review & Editing. **Nigel Jepson:** Conceptualization, Writing - Review & Editing. **Claudio Chiastra**: Methodology, Writing - Review & Editing. **Susann Beier**: Conceptualization, Writing - Review & Editing, Supervision, Project administration.

## Abstract

**Background**

Double Kissing (DK) Crush is a two-stent technique for complex coronary bifurcation lesions, yet the biomechanical influence of plaque on its performance remains poorly understood. This study developed a computational biomechanical model of the DK-Crush procedure to quantify how plaque presence and composition affect procedural outcomes and the performance of Xience Sierra and Orsiro stents.

**Methods**

A population-representative coronary bifurcation was modelled with no plaque, lipid plaque, and fibrous plaque. The complete DK-Crush sequence was simulated using finite element analysis for both stent platforms. Mechanical outcomes included arterial wall stress, malapposition, side branch ostium clearance, and residual stenosis. Post-deployment hemodynamics was assessed using pulsatile computational fluid dynamics, quantifying high shear rate volume and lumen area exposed to low time-averaged endothelial shear stress (TAESS).

**Results**

Plaque presence and stiffness reduced lumen restoration, increased arterial wall stress, led to larger high shear rate regions and, for fibrous plaque, increased exposure to low TAESS. Malapposition and ostial clearance depended mainly on stent design. Plaque also altered the relative performance of the two platforms, revealing differences not observed in plaque-free models.

**Conclusions**

Plaque characteristics substantially affect DK-Crush biomechanics and modify stent behaviour. Incorporating plaque is therefore essential for realistic computational evaluation of bifurcation stenting.

- **First biomechanical study** to evaluate how plaque presence and composition influence biomechanical outcomes of double kissing crush stenting.
- **Plaque characteristics impacted key outcome metrics,** particularly residual stenosis, arterial stresses, and flow, while malapposition and ostium clearance remained mostly driven by stent design.
- **Plaque characteristics are more important than stent platform selection in DK-Crush,** with some differences between Xience Sierra and Orsiro preserved and others changing once plaque was included.

1. **Introduction**

Coronary artery disease is a leading cause of mortality worldwide and frequently requires revascularisation to restore blood flow in narrowed coronary vessels [1]. A considerable proportion of these lesions occur at bifurcations, where a Main Vessel (MV) divides into two branches, the Distal Vessel (DV) and Side Branch (SB). Bifurcation lesions account for about 15-20% of all coronary artery disease patients [2] and are clinically important because their branching geometry favours disturbed flow patterns and plaque development, and complicates percutaneous coronary intervention [3-5]. These anatomical features also make stent deployment more difficult than in straight segments, contributing to higher rates of incomplete scaffolding, malapposition, and adverse events [6].

For complex bifurcation lesions with significant SB involvement, the Double Kissing Crush (DK-Crush) technique is one of the most robust data-driven two-stent strategies [7-9]. Conceptually, DK-Crush begins by placing a stent in the SB, after which the stent is crushed from the MV to ensure complete coverage of the ostium. The technique then uses two rounds of simultaneous inflation of balloons in both branches, known as Kissing Balloon Inflation (KBI), to remove crushed struts from the SB opening and restore a wide, unobstructed ostium [10]. Despite its overall procedural success [11], DK-Crush remains technically demanding and can lead to adverse outcomes associated with In-Stent Restenosis (ISR) and Thrombosis (IST) [12].

Computational biomechanical modelling has become a valuable tool for elucidating these mechanisms, enabling precise quantification of lumen expansion, arterial wall stress, and post-procedural hemodynamic metrics that are difficult or impossible to measure in vivo [7]. Prior biomechanical studies have investigated individual aspects of DK-Crush, including the influence of the Proximal Optimisation Technique (POT) [13], the positioning of the optimisation balloon [14], rewiring strategy [15], and stent platform [16]. Although some of these studies incorporated plaque to reproduce patient-specific geometry [13, 14], they did not analyse how plaque morphology and composition influence deployment mechanics or the resulting flow environment. Moreover, only a limited number of biomechanical models of DK-Crush are currently available [15, 16], and they represent an early stage in the systematic analysis of this technique.

Plaque is often omitted in computational studies of bifurcation stenting because its inclusion increases pre-processing effort, complicates mesh generation, and substantially prolongs simulations [7, 17, 18]. Nonetheless, clinical imaging reveals considerable variability in plaque composition, eccentricity, and stiffness across patients [19], and computational investigations have demonstrated that plaques can restrict lumen expansion, impair scaffolding, and increase arterial stresses [20-25]. Residual stenosis after stenting has also been shown to alter the flow environment by expanding regions of low Time-Averaged Endothelial Shear Stress (TAESS) [22, 26, 27]. Clinical evidence reinforces these insights, since residual stenosis has been linked to higher rates of ISR when greater than 20% [28] and to increased IST [29]. These observations highlight the need to understand how plaque presence and material properties affect the combined mechanical and hemodynamic performance of DK-Crush. In addition, it is unclear how plaque characteristics can alter the performance of current-generation stent platforms.

This study aimed to quantify the influence of plaque presence and composition on the biomechanical performance of the DK-Crush technique. An idealised bifurcation model was used to isolate plaque-related effects and enable controlled comparisons. Within this framework, the effects of lipid and fibrous plaques on lumen restoration, malapposition, arterial wall stress, and flow patterns were evaluated, and their impact on the performance of two contemporary stent platforms was assessed.

## 2. Materials and Methods

### 2.1. Deployment Mechanics Simulation

A population-representative coronary bifurcation model was generated using Finet's law to set the MV, DV, and SB diameters to 3.25 mm, 2.5 mm, and 2.3 mm [30], with a bifurcation angle of 70° and uniform wall thickness of 0.89 mm [31] (Figure 1). The arterial wall consisted of intima, media, and adventitia layers. Two plaque geometries were included, one along the MV-DV segment and the other within the SB, since disease extending across all three branches represents the most common clinical case, affecting 36% of patients [32] (Figure 1). Plaque morphology varies substantially in vivo, so two common patterns were selected. The SB plaque was concentric with a 40% stenosis, whereas the MV plaque was eccentric with a half-moon profile and a 20% stenosis. The lower stenosis severity in the MV reflects the residual narrowing typically present after lesion preparation. In clinical practice, balloon dilatation of lipidic lesions and scoring-balloon modification of fibrous tissue reduce the obstruction before stent delivery [5]. The chosen level of stenosis therefore represents a post-preparation lesion that still affects expansion mechanics but does not impede device passage.

To reproduce the nonlinear response of vascular tissue under large deformation, all arterial layers and plaques were modelled as isotropic hyperelastic materials using a reduced-polynomial strain-energy density function [33]:

$$U = \sum_{i=1}^{n} C_{i0}(\bar{I}_1 - 3)^i$$

where $C_{i0}$ are the material coefficients reported in Table 1 and $\bar{I}_1$ is the first invariant of the deviatoric deformation tensor. Both plaque hyperelastic material models were coupled with perfect plasticity to model rupture [34-36] (Figure 1). The bifurcation geometry with plaques was meshed with 632,592 C3D8R hexahedral elements, containing 9 elements across the artery wall thickness and 6 elements across the plaque thickness [16].

Stent geometries for Xience Sierra (Abbott Vascular, Abbott Park, IL, USA) and Orsiro (Biotronik AG, Bulach, Switzerland) were reconstructed from micro-CT data and parametrically rebuilt [37-39] (Figure 1). Both stent platforms were made of a cobalt–chromium alloy that was modelled with an elasto-plastic constitutive model, with parameters in Table 1 [16]. Stent geometries were meshed with 98,428-197,060 C3D8R hexahedral elements, depending on stent length and pattern, with 4 elements through the strut thickness [16]. Angioplasty balloons were modelled following an established approach [40]. Balloon geometries with the required diameters and lengths were generated to match the devices used during each step of the DK-Crush sequence, based on the Accuforce balloon (Terumo, Japan), which is widely used in bifurcation interventions due to its low compliance. Each balloon was discretised with M3D4R elements with a thickness of 30 μm. Depending on balloon size, the mesh contained between 4320-6360 elements. The polymer material was modelled as isotropic linear elastic with density set to 1000 kg/m$^3$ and Poisson's ratio to 0.45. The Young's modulus was calibrated for each balloon size by matching the simulated pressure-diameter response to the manufacturer compliance charts, ensuring that the expansion behaviour of the numerical balloons reproduced that of the Accuforce device.

Table 1 – Material coefficients used in the finite element simulations for the arterial layers, lipid and fibrous plaques, and stent [33, 35].

|  | Density | $C_{10}$ | $C_{20}$ | $C_{30}$ | $C_{40}$ | $C_{50}$ | $C_{60}$ | Yield Stress |
|---|---|---|---|---|---|---|---|---|
|  | [kg/m$^3$] | [MPa] | | | | | | |
| Intima |  | 6.79E-03 | 5.40E-01 | -1.11 | 10.65 | -7.27 | 1.63 | - |
| Media |  | 6.52E-03 | 4.89E-02 | 9.26E-03 | 7.6E-01 | -4.3E-01 | 8.69E-02 | - |
| Adventitia | 1120 | 8.27E-03 | 1.20E-02 | 5.20E-01 | -5.63 | 21.44 | 0 | - |
| Lipid plaque |  | 4.50E-02 | 1.7E-01 | -1.3E-01 | 1.1E-01 | - | - | 0.19 |
| Fibrous plaque |  | 1.0E-02 | 4.9E-01 | 4.13 | - | - | - | 2.07 |
|  | Density | Young Modulus | Yield Stress | Ultimate Stress | Ultimate deformation | Poisson's coeff. | | |
|  | [kg/m$^3$] | [MPa] | | | - | - | | |
| Stent alloy | 8000 | 233,000 | 414 | 933 | 45% | 0.35 | | |

DK-Crush deployment was simulated in Abaqus/Explicit (Dassault Systèmes, Providence, RI, USA). The complete clinical workflow was reproduced (Figure 2), including (1) SB stent crimping and deployment, (2) crush, (3) POT crush, (4) proximal rewiring, (5) first KBI, (6) MV stent crimping and deployment, (7) first POT, (8) proximal rewiring, (9) second KBI, and (10) final POT, following the protocol used in previous work [16]. Each balloon inflation was driven by uniform internal pressure, followed by a short unloading phase to allow elastic recoil [16]. Vessel ends were fixed to prevent rigid body motion. Interactions among vessel, stents, and balloons were handled using a general contact formulation with hard normal contact and a friction coefficient of 0.2 [33, 41]. Semi-automatic mass scaling and smoothed loading curves were used to ensure quasi-static conditions, verified by maintaining the kinetic-to-internal energy ratio below 5% [42]. All deployment simulations were executed in parallel on two compute nodes (96 cores total) of the Gadi high-performance computing system (NCI Australia). Each node was equipped with dual Intel Xeon Platinum 8274 Cascade Lake CPUs at 3.2 GHz and 192 GB RAM.

Four outcomes were extracted from the final deployment configuration:

- Arterial wall stress. Elevated wall stress has been associated with localised injury, inflammation, and pathways contributing to ISR [43]. Two related quantities were evaluated: (i) the percentage of arterial volume with stress exceeding 10 kPa, and (ii) the average maximum principal stress within that region [16].
- Malapposition. Defined as the percentage of stent surface located more than one strut thickness away from the arterial wall. Malapposed struts impair endothelial healing and have been linked to an increased risk of IST [44].
- SB ostium clearance. Quantified as the percentage of the SB ostium area free from strut obstruction. Adequate ostial clearance facilitates access for future re-access and reduces the likelihood of flow disturbance at the bifurcation carina [45].
- Residual stenosis. Represents the degree of lumen narrowing that remains after DK-Crush, calculated as the percentage reduction in post-stenting lumen size relative to the healthy reference diameter. Residual stenosis has been clinically associated with higher rates of ISR and IST, and can exacerbate post-procedural flow disturbances, making it a key indicator of scaffolding performance [28, 29].

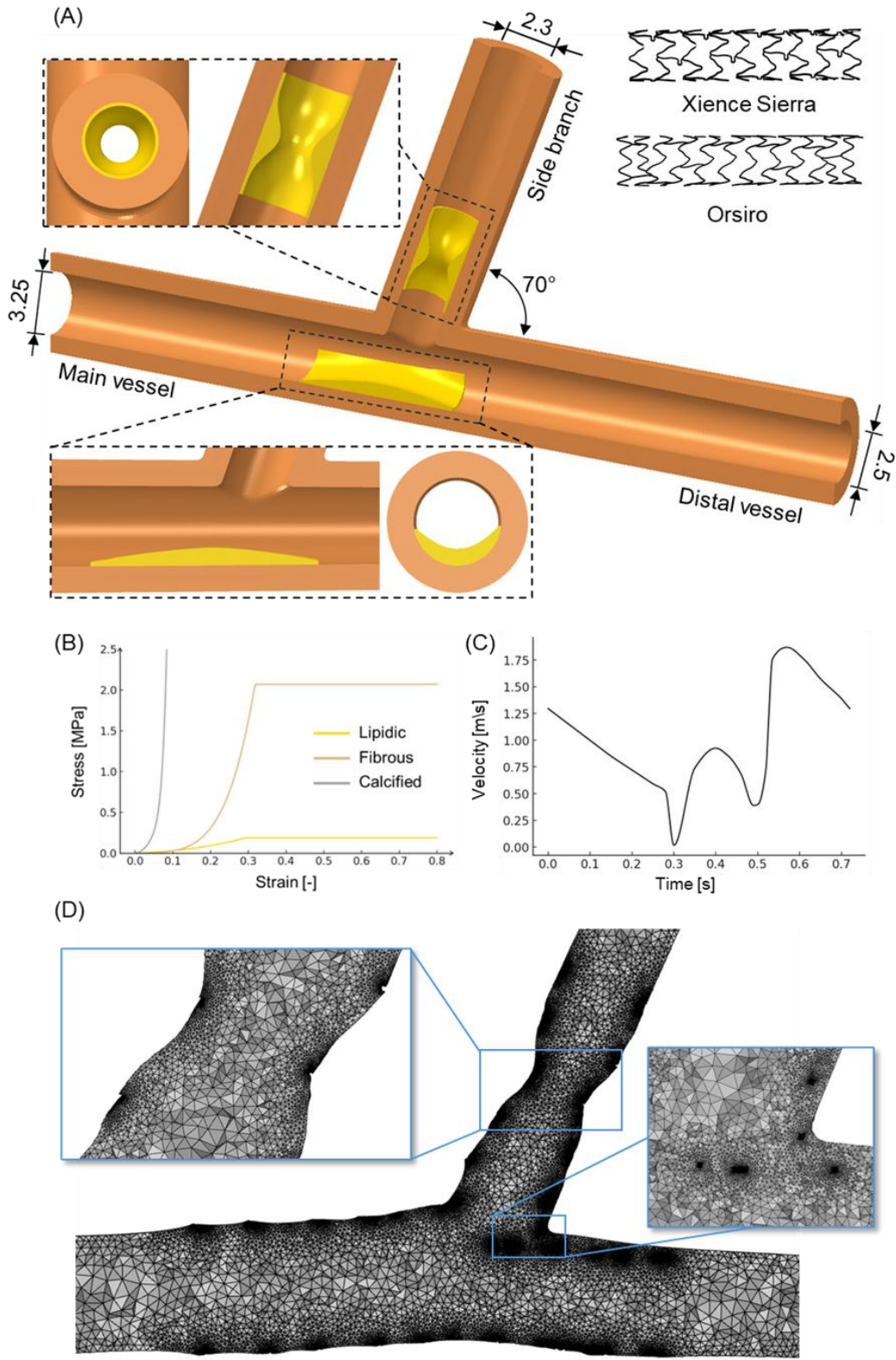

**Fig. 1** (A) Generalised coronary bifurcation model with a 70° bifurcation angle. A concentric plaque was placed in the SB and an eccentric plaque along the MV-DV segment. The two commercial stents modelled were Xience Sierra (81 µm) and Orsiro (60 µm). (B) Stress-strain curves for the two plaque material models. Lipid and fibrous properties were used in the simulations, each coupled with a plasticity model [35], while the calcified response is shown only for reference and comparison. (C) Physiological inlet velocity waveform applied at the MV inlet to reproduce coronary pulsatile flow. (D) Tetrahedral CFD mesh of the post-deployment lumen, with local refinement around the plaques and stented region. CFD: Computational Fluid Dynamics; MV: Main Vessel; SB: Side Branch

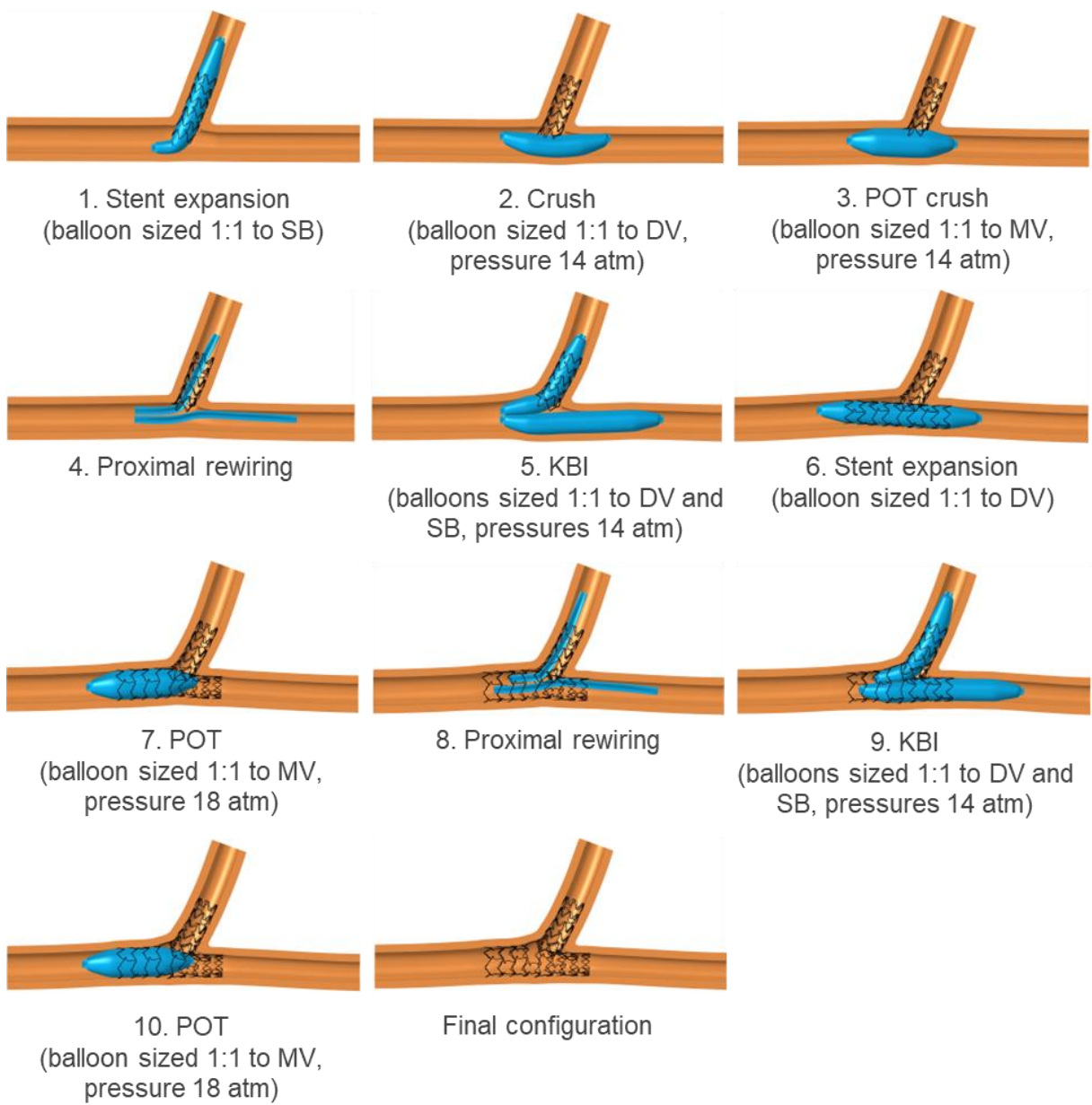

**Fig. 2** Procedural steps followed in the FEA to replicate the DK-Crush coronary bifurcation stenting technique. Image adapted with permission from [16]. *DK: Double Kissing; DV: Distal Vessel; FEA: Finite Element Analysis; KBI: Kissing Balloon Inflation; MV: Main Vessel; POT: Proximal Optimisation Technique; SB: Side Branch*

## 2.2. Computational Fluid Dynamics Simulations

The lumen geometry after complete DK-Crush deployment was exported from the FEA and discretised with tetrahedral elements using ANSYS Fluent Meshing (ANSYS, Inc., Canonsburg, PA, USA). Mesh refinement was concentrated near the stent, with a minimum element size of 14 µm (Figure 1). The final meshes contained between 9 and 10 million elements. A mesh independence test confirmed stable wall shear metrics beyond nine million elements.

Blood was modelled as an incompressible non-Newtonian fluid governed by the Navier-Stokes equations, with viscosity defined by the Carreau-Yasuda model [46]. A constant density of 1050 kg/m³ was used [46]. All arterial walls were assumed rigid with a no-slip boundary condition [46, 47]. A time-varying parabolic velocity profile was imposed at the MV inlet to reproduce the pulsatile nature of coronary blood flow (Figure 1). The waveform was derived from a representative physiological velocity curve [46] and scaled to the vessel size using an allometric relationship between diameter and flow rate [48, 49]:

$$Q_{MV} = 1.43 d_{MV}^{2.55}$$

Where $d_{MV}$ is the mean diameter of the MV. At the outlets, the flow division between DV and SB was determined from [49]:

$$\frac{Q_{DV}}{Q_{SB}} = \left(\frac{d_{DV}}{d_{SB}}\right)^{2.27}$$

Where $Q_{DV}$ and $Q_{SB}$ are the outlet flow rates and $d_{DV}$ and $d_{SB}$ the mean diameters. The flow split was calculated as 54.7% through DV and 45.3% through SB.

Simulations were performed in ANSYS CFX using a second-order backward-Euler scheme, while spatial advection was treated with the high-resolution (bounded second-order upwind) scheme.

Convergence criterion for velocity and continuity residuals was set to $10^{-5}$ [4]. Each case was run for four cardiac cycles at a time step of 0.001 s, with results extracted from the last cycle to minimise transient start-up effects [4]. All CFD simulations used 48 cores on Gadi.

Two CFD metrics were extracted:

- Shear rates: a strong indicator of local platelet activation, since regions where shear rates exceed 1000 s⁻¹ are considered thrombogenic [50]. Two related quantities were extracted at the time of peak inlet velocity i) the volume of blood exposed to shear rate above 1000 s⁻¹, and (ii) the average shear rate within that volume, to characterise the extension and intensity of thrombogenic flow. This time point was selected because shear rates reach their maximum at peak inflow, which makes it the phase most representative of the highest thrombogenic potential.

- The percentage of arterial surface exposed to TAESS < 0.4 Pa, which is associated with neointimal hyperplasia, restenosis, and impaired endothelial function [51].

## 3. Results

### 3.1. Mechanical Outcomes

The full FEA DK-Crush simulation without plaque required approximately 5 hours of wall-clock time. The plaque cases required substantially longer due to increased stiffness, additional contact interactions, and mass scaling, with wall-clock times ranging from 28 to 40 hours.

In the no-plaque configuration, Xience Sierra and Orsiro exhibited comparable average maximum principal stresses (33 vs. 35 kPa), with identical proportions of arterial tissue exposed to stresses exceeding 10 kPa (16.0%) (Figure 3). With lipid plaque, both stents showed higher artery stresses, which increased to 52 kPa for Xience Sierra and 43 kPa for Orsiro, while the high-stress volume rose to 44% and 34%, respectively. In the fibrous plaque case, stresses increased further, reaching 60 kPa for Xience Sierra and 44 kPa for Orsiro, with high-stress volumes of 44% and 31%. In all scenarios, the peak artery wall stresses were located at the SB ostium opposite the carina.

Malapposition rates were consistently higher for Orsiro than for Xience Sierra in all scenarios (Figure 4). Malapposition in the no-plaque model was 12.2% for Xience Sierra and 15.3% for Orsiro. In the lipid plaque case, malapposition increased to 13.6% and 19.9%, respectively. In the fibrous plaque case, malapposition was 14.4% for Xience Sierra and 19.2% for Orsiro.

SB ostium clearance was identical between Xience Sierra and Orsiro in the no-plaque model (52%) (Figure 5). With lipid plaque, ostium clearance was 53% for Xience Sierra and 45% for Orsiro. In the fibrous plaque case, Xience Sierra and Orsiro achieved 45% and 47% ostium clearance respectively.

Residual stenosis increased with plaque presence and was consistently higher for Orsiro (Figure 6). Residual stenosis in the lipid plaque model reached 13.7% for Xience Sierra and 18.0% for Orsiro, while in the fibrous plaque model values increased to 26.4% and 30.6%, respectively.

### 3.2. Hemodynamic Outcomes

Each CFD simulation required between 3 and 6 hours of wall-clock time, with longer runtimes observed for the models with plaque, due to their increased mesh sizes.

Average shear rate at peak systole was similar across all configurations for both stent platforms (Xience Sierra: 1472-1543 $s^{-1}$; Orsiro: 1463-1616 $s^{-1}$) (Figure 7), indicating no substantial differences between stents or plaque types in this metric. In contrast, the volume exposed to high shear rate (> 1000 $s^{-1}$) increased markedly in the presence of plaque, with a larger rise as residual stenosis became more severe. In the no-plaque model, these volumes were 0.08 mm³ for Xience Sierra and 0.10 mm³ for Orsiro. They increased to 0.17 mm³ and 0.40 mm³ in the lipid plaque case, and to 0.89 mm³ and 1.15 mm³ in the fibrous plaque case, respectively.

The percentage of the lumen area exposed to TAESS below 0.4 Pa also varied with plaque types and demonstrated a shift in the relative difference between platforms (Figure 8). In the no-plaque configuration, Xience Sierra had a higher low-TAESS area than Orsiro (36.3% vs. 28.2%). With lipid plaque, the percentage of low-TAESS area decreased and became similar between stents (32.3% vs. 31.6%). In the fibrous plaque configuration, both platforms exhibited similarly elevated low-TAESS lumen area (38.7% vs. 40.0%).

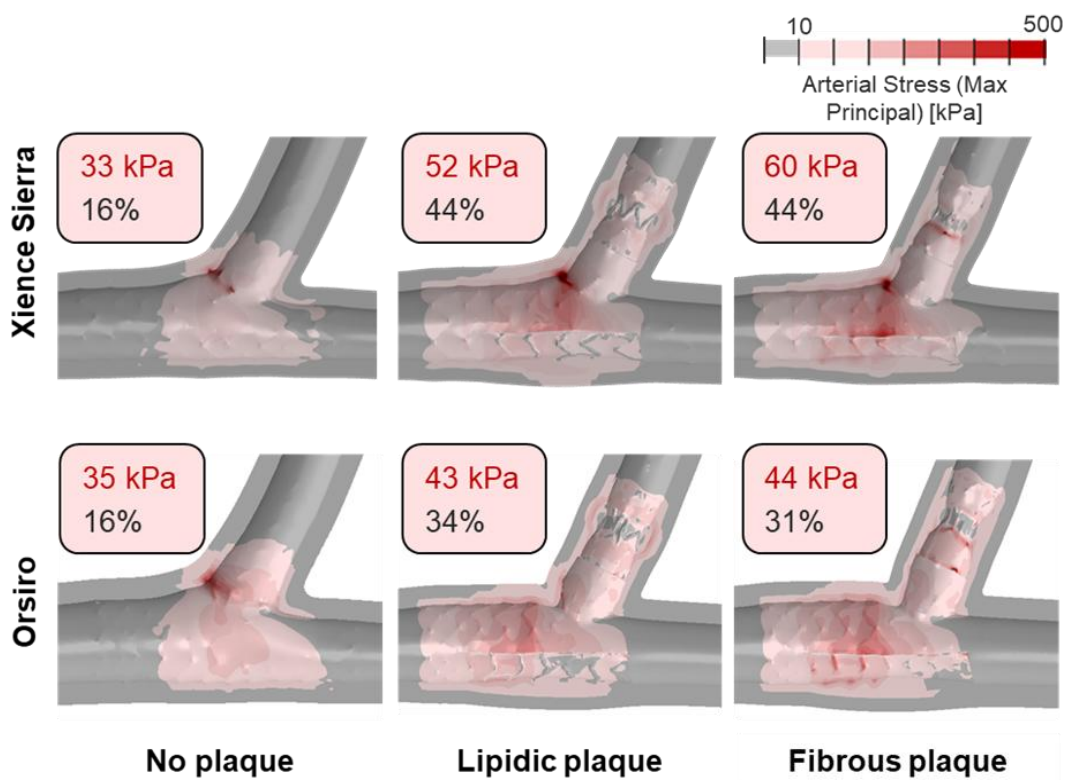

**Fig. 3** Arterial wall stress distribution (maximum principal stress) after double-kissing crush stenting for Xience Sierra (top) and Orsiro (bottom) under different plaque conditions. The color scale represents stress magnitude, while the insets indicate the average wall stress and percentage of vessel area exceeding 10 kPa

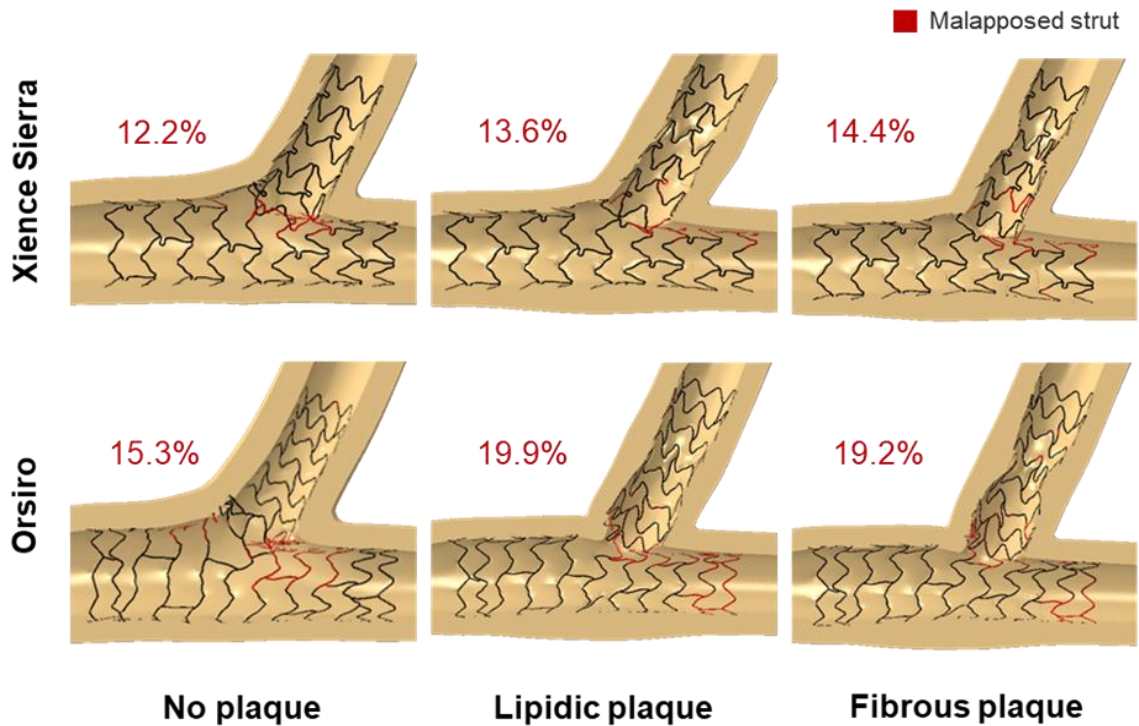

**Fig. 4** Malapposition distribution after double-kissing crush stenting for Xience Sierra (top) and Orsiro (bottom) under three conditions: no plaque, lipidic plaque, and fibrotic plaque. Red struts indicate malapposed regions, and the percentage values represent total malapposition area relative to vessel surface

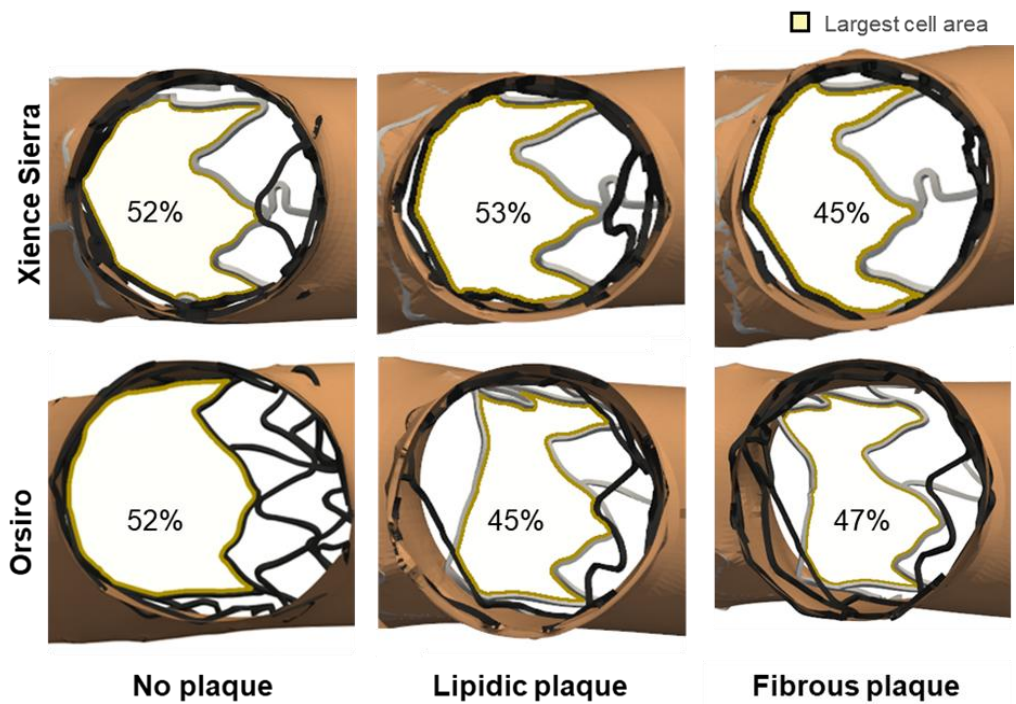

**Fig. 5** Cross-sectional views of the side branch ostium showing the largest accessible cell area after double-kissing crush stenting for Xience Sierra (top) and Orsiro (bottom) under different plaque conditions. The percentage values indicate the side branch ostium clearance relative to the vessel area

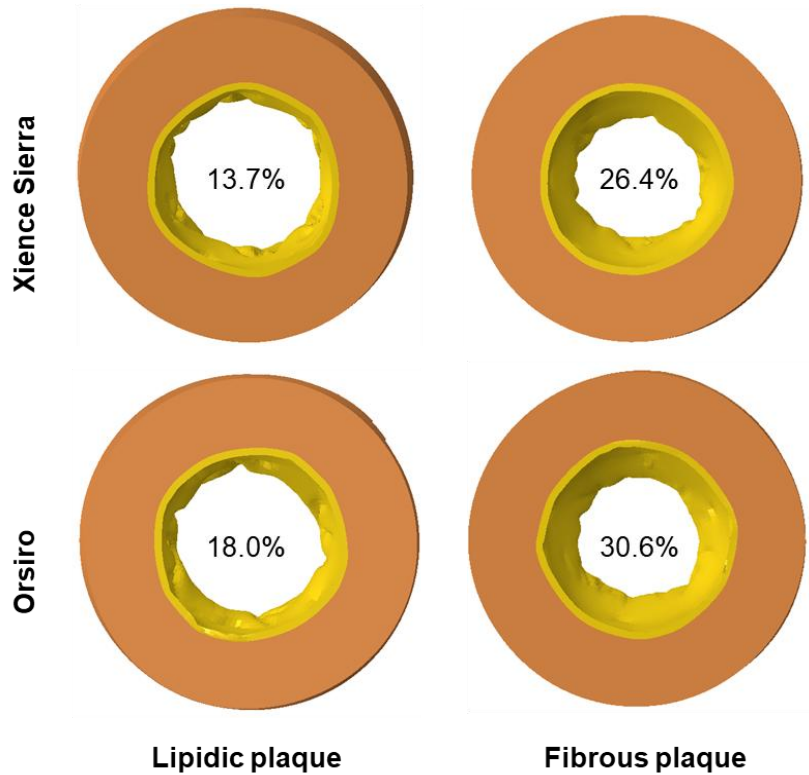

**Fig. 6** Cross-sectional views showing SB residual stenosis after double-kissing crush stenting for Xience Sierra (top) and Orsiro (bottom) under different plaque conditions. The percentage values indicate residual stenosis relative to the healthy reference lumen. *SB: Side Branch*

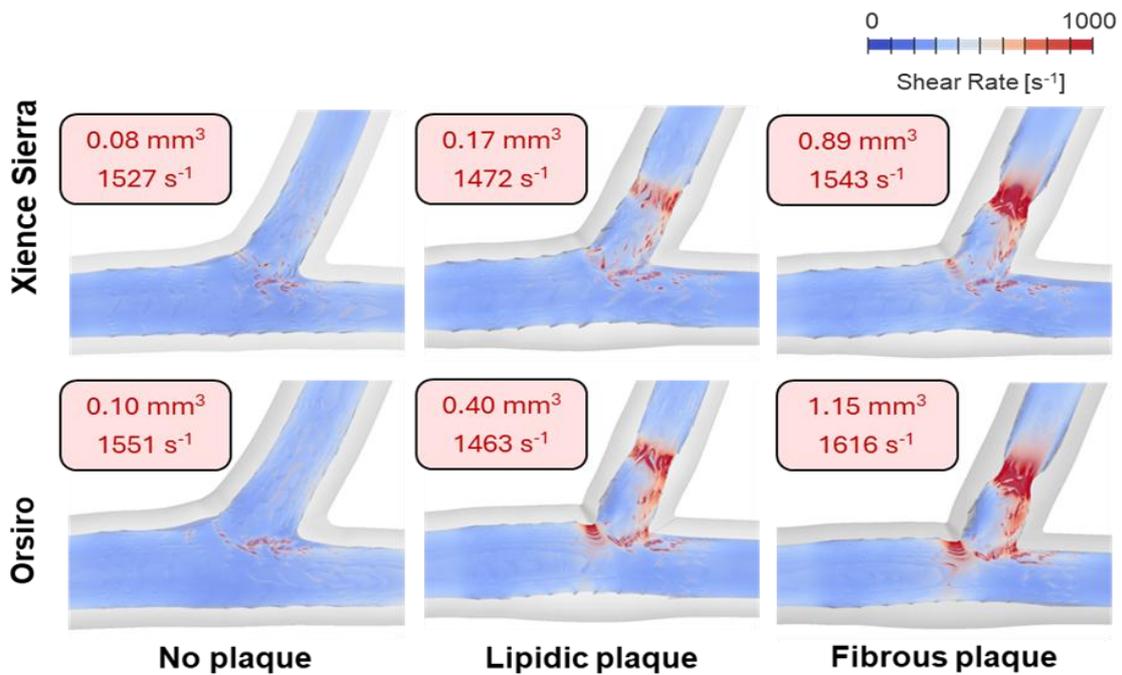

**Fig. 7** Shear rate distribution after double-kissing crush stenting for Xience Sierra (top) and Orsiro (bottom) under different plaque conditions. The color scale represents instantaneous shear rate magnitude at maximum inlet velocity, and the insets indicate the volume of regions exceeding 1000 s$^{-1}$ and the corresponding average shear rate

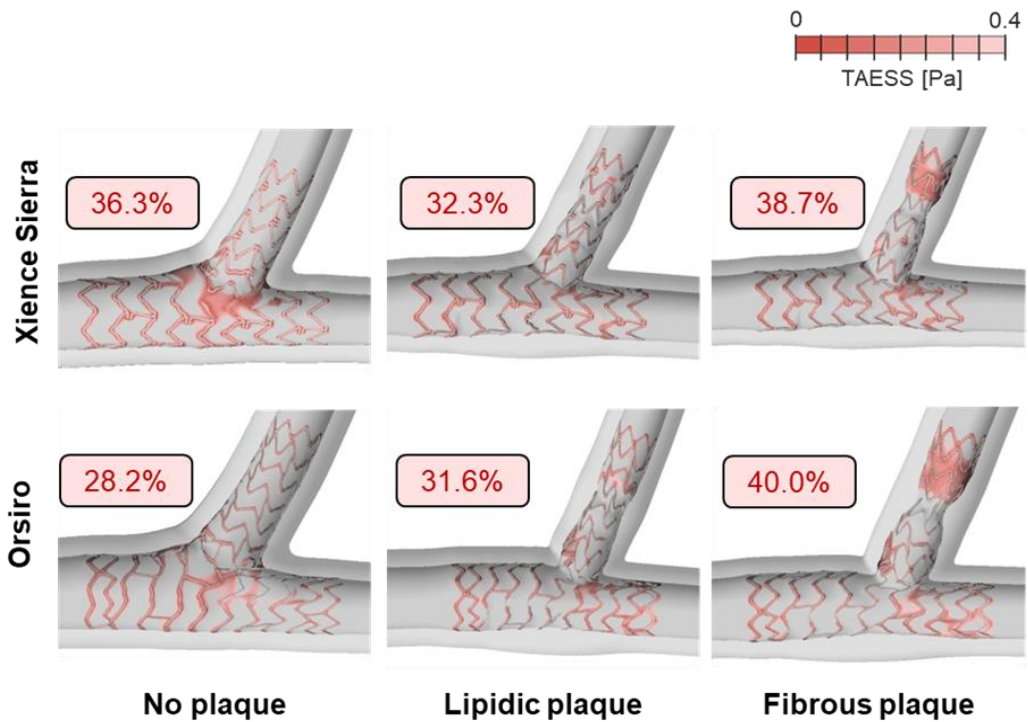

**Fig. 8** Distribution of TAESS after double-kissing crush stenting for Xience Sierra (top) and Orsiro (bottom) under different plaque conditions. The color scale represents TAESS magnitude, and the percentage values indicate the vessel area exposed to low shear stress (< 0.4 Pa). *TAESS: Time-Averaged Endothelial Shear Stress*

## 4. Discussion

This study presents an in silico biomechanical model of the DK-Crush coronary stenting procedure, developed to investigate how disease characteristics shape mechanical expansion and post-procedural flow. Within this framework, we evaluated the influence of plaque by comparing plaque-free, lipidic, and fibrous configurations. The results show that plaque presence and composition affect lumen restoration, arterial wall stress, flow patterns, and the relative performance of different stent platforms. Incorporating lipid and fibrous plaques revealed interactions that do not appear in plaque-free simulations, highlighting the value of accounting for representative disease features when conducting detailed computational biomechanical assessments of complex two-stent techniques such as DK-Crush.

### 4.1. Impact of Plaque on DK-Crush and Stent Performance

Plaque presence and material properties altered several aspects of the biomechanical performance of DK-Crush and the behaviour of both contemporary stent platforms evaluated. Both lipid and fibrous plaques reduced scaffolding effectiveness and increased arterial wall stress, with more pronounced effects for the stiffer fibrous lesion.

Arterial wall stress exhibited a strong dependence on plaque material properties, consistent with previous studies [20, 22, 23]. Both lipid and fibrous plaques led to higher average stresses and enlarged regions of elevated stress within the bifurcation. The two stents performed similarly in the no-plaque model, but clear differences emerged once plaque was included. In both plaque scenarios, Xience Sierra imparted higher stresses to the bifurcation wall than Orsiro, indicating that incorporating plaque allows computational models to better capture platform-specific behaviour that may not be apparent in idealised plaque-free simulations.

Malapposition showed a smaller sensitivity to lesion characteristics. Values increased slightly with both plaque types, reflecting localised alterations in contact as the vessel expanded against a stiffened lesion. Orsiro consistently showed higher malapposition than Xience Sierra, consistent with its thinner struts providing less structural support. Although plaque contributed to a modest rise in malapposition, overall patterns remained primarily stent-driven.

SB ostium clearance also showed limited sensitivity to plaque. Differences between plaque types were small for both platforms. Instead, ostium clearance appeared mainly governed by stent positioning and rewiring configuration, which remains central determinants of ostial reopening in DK-Crush [15, 16].

Residual stenosis was strongly influenced by plaque characteristics. Both lipid and fibrous plaques generated considerable post-procedural narrowing, with the fibrous plaque resulting in the largest reductions in lumen size. Between the two stents, Orsiro consistently exhibited higher residual stenosis than Xience Sierra, reflecting its thinner design and reduced radial support. These results indicate that scaffolding quality depends strongly on the lesion stiffness and on stent mechanical performance, with plaque playing a dominant role in determining the degree of lumen restoration achievable with DK-Crush.

Shear rate behaviour followed directly from these mechanical differences. The volume of blood exposed to high shear rates increased with higher residual stenosis. Since residual stenosis was shown to be associated with increased rates of IST [29] and high shear rates are known to promote platelet activation [50], our results may explain the clinical association between residual stenosis and IST. Across cases, Orsiro produced larger high-shear volumes than Xience Sierra, in line with its higher malapposition and reduced scaffolding capacity.

Low-TAESS exposure followed a distinct pattern. Small changes occurred in the lipid plaque case compared to the no-plaque scenario. Differently, the fibrous plaque produced a marked increase in low-TAESS regions downstream of the stented plaque. This agrees with clinical data showing increased ISR rates when residual stenosis exceeds 20% [28], a threshold surpassed in our simulations only by the fibrous plaque configuration. The agreement between these patterns and clinical observations provides a clear mechanistic link between lumen reduction and the hemodynamic environments known to promote ISR.

Overall, this analysis demonstrates that plaque and stent design influenced DK-Crush outcomes to different extents depending on the metric considered. Malapposition and ostial clearance were shaped mainly by stent configuration and procedural technique and changed only modestly with plaque. In contrast, arterial wall stress, residual stenosis, and the associated hemodynamic indices were dominated by plaque characteristics, showing interactions between lesion stiffness and stent behaviour. These findings indicate that plaque burden may be a more influential determinant of DK-Crush performance than stent platform selection for several clinically relevant metrics.

### 4.2. Modeling Implications

The findings of this study carry several implications for computational modelling of DK-Crush. Plaque characteristics strongly influence several key mechanical and hemodynamic outcomes, particularly residual stenosis, arterial wall stress, and flow patterns. Models that omit plaque are therefore likely to underestimate vessel loading and overestimate the degree of flow restoration. This is consistent with prior work in other stenting contexts, where lesion characteristics were shown to affect expansion behaviour and local hemodynamics [17, 20-27]. In the setting of DK-Crush, the present results reinforce that disease characteristics may shape outcomes as strongly as stent design.

The substantial variability of plaque morphology across patients presents a challenge for accurate modelling. Real lesions differ markedly in composition, thickness, and eccentricity, and no single idealised model can capture this clinical diversity [17]. Because lumen restoration and hemodynamics were sensitive to plaque stiffness in this study, imaging-derived plaque morphology and material properties are essential when the objective is to generate clinically meaningful predictions, such as estimating the likelihood of ISR or IST or evaluating how a specific lesion interacts with device choice or procedural strategy [52].

Plaque-free models nevertheless serve an important complementary role. By avoiding disease-related variability and requiring significantly lower computational cost, they facilitate extensive parametric studies, device design screening, and early-phase testing. In these scenarios, the objective is often to identify stent-or procedural-driven differences rather than to predict absolute clinical values. Simplified geometries are well suited to this purpose and can be used as an efficient and informative foundation before introducing plaque when lesion-specific interactions become a central focus [7].

Overall, both approaches offer distinct strengths. Simplified geometries provide a fast, reproducible, and generalisable basis for identifying device- or procedure-related differences, whereas plaque-inclusive simulations deliver the fidelity needed when lesion characteristics are integral to the question being addressed. Combining these approaches enables broader generalisability in idealised studies and greater accuracy in patient-specific or preclinical planning tasks.

### 4.3. Limitations

First, the bifurcation anatomy and plaque geometries did not fully capture the anatomical diversity and lesion complexity observed clinically. Real plaques often exhibit heterogeneous composition, irregular shapes, and calcification, all of which may further influence procedural outcomes. In this analysis, only one plaque configuration was examined and only the material properties were varied, which provides controlled comparisons but does not reproduce the full spectrum of disease encountered in practice. In addition, the simulations did not include lesion preparation. Clinically, lipidic plaques are typically pre-dilated and fibrous lesions are modified with scoring or cutting balloons before stent implantation, which reduces the obstruction and facilitates device delivery. These steps can alter the mechanical resistance of the lesion, produce a pre-stress field in the plaque, and may affect expansion behaviour relative to the untreated plaques modelled here. Second, the analysis was limited to a single stenting technique and two contemporary stent platforms. Although DK-Crush is widely used for treating complex bifurcations, the stent-plaque interactions observed in this study may differ for other devices or techniques. Additional work is needed to determine whether the trends identified here apply to other stent designs or bifurcation strategies. Finally, the hemodynamic simulations assumed rigid vessel walls and did not include cardiac motion, vessel compliance, or long-term biological processes. These factors can gradually modify flow patterns and mechanical loading after implantation. Multiscale and systems biology models have been used to simulate long-term vascular adaptation and restenosis progression [53, 54]. Incorporating these elements in future work would provide a more comprehensive assessment of DK-Crush performance under pathophysiological conditions.

This study represents the first biomechanical evaluation of the DK-Crush technique that incorporates plaque to assess how plaque characteristics influence procedural outcomes. Plaque presence and material properties substantially modified several DK-Crush metrics, particularly scaffolding, arterial wall stress, and hemodynamic indices, whereas others, such as malapposition and ostial clearance, remained mainly driven by stent design. Plaque characteristics also altered the relative performance of the two stent platforms, outweighing some device-specific differences and revealing interactions not captured in plaque-free simulations. Overall, these findings underscore the importance of representing plaque in computational models of complex stenting techniques such as DK-Crush and constitute an initial step toward incorporating greater lesion complexity into future biomechanical assessment.

**Acknowledgments**

This study includes computations performed on the computational cluster Katana (supported by Research Technology Services at UNSW Sydney) and from the computational cluster Gadi (supported by National Computational Infrastructure, an NCRIS-enabled facility supported by the Australian Government).